\documentclass[twoside]{article}
\usepackage{fleqn,espcrc2,epsfig}
\title{Analysis of oscillations of atmospheric neutrinos}

\author{G.L.\ Fogli\address{Dipartimento di Fisica and Sezione INFN di Bari,
	Via Amendola 173, 70126 Bari, Italy},
	E.\ Lisi$^{\rm a}$\thanks{Speaker. E-mail: \tt lisi@ba.infn.it},
	A.\ Marrone$^{\rm a}$, and 
	D.\ Montanino\address{Dipartimento di Scienza dei Materiali
	dell'Universit\`a di Lecce, Via Arnesano, 73100 Lecce, Italy}}
\begin{document}
\begin{abstract}
We briefly review  the current status of standard oscillations of atmospheric
neutrinos in schemes with two, three, and four flavor mixing. It is shown that,
although the pure $\nu_\mu\to\nu_\tau$ channel provides an excellent $2\nu$ fit
to the data, one cannot exclude, at present, the occurrence of additional 
subleading $\nu_\mu\to\nu_e$ oscillations ($3\nu$ schemes) or of sizable
$\nu_\mu\to\nu_s$ oscillations ($4\nu$ schemes).  It is also shown that the
wide dynamical range of energy and pathlength probed by the  Super-Kamiokande
experiment puts severe constraints on  nonstandard explanations of the
atmospheric neutrino data, with a  few notable exceptions.
\vspace{1pc}
\end{abstract}
\maketitle

\section{Introduction}

It is well known that the Super-Kamiokande (SK) atmospheric neutrino data can
be beautifully explained in terms of $2\nu$ oscillations in the
$\nu_\mu\to\nu_\tau$ channel \cite{SK00}. This interpretation is also supported
by MACRO \cite{MA00} and by Soudan2 \cite{So00}. Conversely, pure
$\nu_\mu\to\nu_e$ oscillations do not provide a good fit to the SK data
\cite{Fo99}, and are independently excluded by the negative $\nu_e$
disappearance searches in the CHOOZ \cite{CH99} and Palo Verde \cite{Palo}
reactors. Pure $\nu_\mu\to\nu_s$ oscillations ($\nu_s$ being a hypothetical
sterile neutrino)  are also disfavored by SK \cite{SK00,SKst} (and by MACRO
\cite{MA00}), due to nonobservation of the associated matter effects
\cite{Matt} and neutral current event depletion \cite{Viss}.

Although two-flavor $\nu_\mu\to\nu_\tau$ oscillations represent the most
economical explanation, it should be stressed that, to some extent, additional
oscillation channels may be open, as naturally expected in $3\nu$ and $4\nu$
schemes \cite{Theo} accommodating the current phenomenology. Moreover,
$\nu_\mu$ disappearance might be driven by dynamics different from the simple
mass-mixing mechanism. In this article, we briefly review the status of such
solutions, with emphasis on: (i) scenarios involving more than two states
($3\nu$ and $4\nu$ mixing), and (ii) scenarios involving nonstandard dynamics
(decay,  extra dimensions, decoherence).

\section{$3\nu$ oscillations}

Assuming that two out of three active $\nu$'s are almost degenerate (say,
$m_1\simeq m_2$), it can be shown \cite{Fo99} that  atmospheric $\nu$'s probe
only $m^2\equiv m^2_3-m^2_{1,2}$ and the mixing matrix elements $U_{\alpha 3}$:
\begin{equation} 3\nu{\rm\ parameter\ space}
\;\equiv\;
(m^2,U^2_{e3},U^2_{\mu 3},U^2_{\tau 3})\ , 
\end{equation} 
with $U^2_{e3}+U^2_{\mu 3}+U^2_{\tau 3}=1$ for unitarity. 
\vspace*{3mm}
\hrule
\begin{figure}[hb]
\vspace*{-1.7truecm}
\hspace*{-0.2truecm}
\epsfig{bbllx=1.4truecm,bblly=1.3truecm,bburx=19.5truecm,bbury=26.5truecm,%
height=10.5truecm,figure=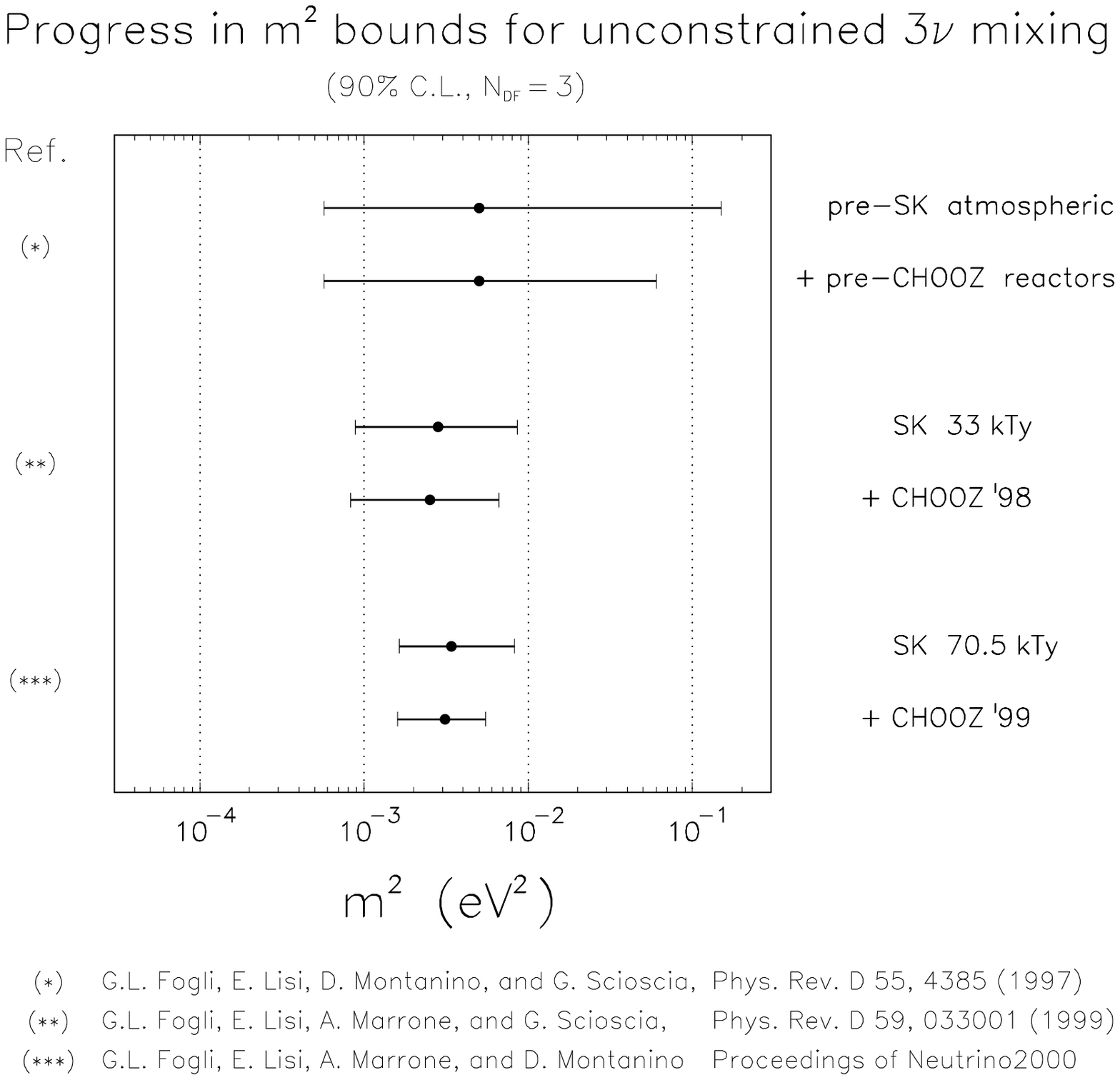}
\vspace*{-4.0cm}
\caption{ Progress in bounds on $m^2$, as derived by $3\nu$ analyses of
atmospheric and reactor data, both before and after SK and CHOOZ.}
\label{f1}
\end{figure}

\begin{figure*}[t]
\vspace*{-4.9truecm}
\hspace*{+.9truecm}
\epsfig{bbllx=1.4truecm,bblly=1.3truecm,bburx=19.5truecm,bbury=26.5truecm,%
height=18.truecm,figure=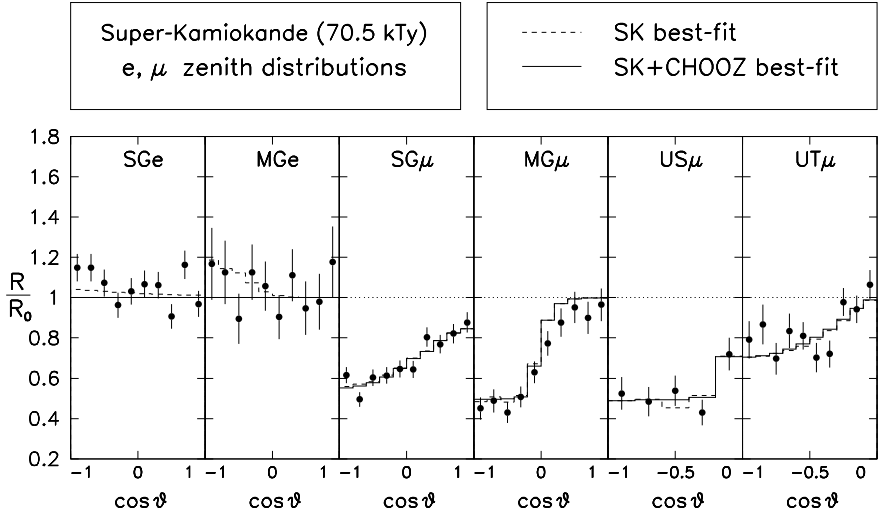}
\vspace*{-7.0cm}
\caption{ SK zenith distributions, normalized to no-oscillation expectations.
Dots with error bars: SK data. Dashed and solid lines: best fits to SK only and
to SK+CHOOZ \protect\cite{Ma00}.}
\label{f2}
\end{figure*}

We present a preliminary update \cite{Ma00} of previous limits \cite{Fo99}
on such parameters, using the latest data from SK (70.5 kTy) \cite{SK00} and
CHOOZ \cite{CH99}. The SK data include 55 zenith  bins: 10+10 bins for the
subGeV (SG) $e$+$\mu$ events, 10+10 bins for the multiGeV (MG) $e$+$\mu$
events, and 5+10 bins for the upward stopping (US) and through-going (UT) $\mu$
events. For CHOOZ, we use the total rate (one datum). We  accurately calculate
all such observables, and $\chi^2$-fit them (see \cite{Fo99} for details).

Figure~1 shows that the allowed range for $m^2$  is stable around $3\times
10^{-3}$ eV$^2$. The same figure also shows the impact of SK and CHOOZ in
sharpening \cite{Ma00,Fo99}  prior bounds on $m^2$ \cite{Fo97}.

Figure~2 shows the SK data and the best-fit theoretical distributions. The best
fit for SK data only ($\chi^2=47.5$, dashed line) is found at 
\begin{equation} (m^2,\,U^2_{e3},\,U^2_{\mu3},\,U^2_{\tau 3})\simeq
(3.5,\,0.07,\,0.57,\,0.36)\ , \end{equation}
where $[m^2]=10^{-3}$ eV$^2$. For $U^2_{e3}=0.07$,  the theoretical  MG$e$
distribution shows a distortion which, however, is well within the
uncertainties. The weak preference for $U^2_{e3}\neq 0$ is  suppressed by CHOOZ
data. The SK+CHOOZ best fit ($\chi^2=49$, solid lines) basically corresponds to
pure $\nu_\mu\to\nu_\tau$ oscillations with maximal mixing,
\begin{equation} (m^2,\,U^2_{e3},\,U^2_{\mu3},\,U^2_{\tau 3})
\simeq (3.0,\,0,\,0.5,\,0.5)\ , \end{equation}
with limited allowance for extra $\nu_e$ mixing \cite{Ma00},
\begin{eqnarray} {\rm SK\ data\ only } &:& U^2_{e3} < 0.31\;\; (0.38)\ ,\\ {\rm
SK + CHOOZ } &:& U^2_{e3} < 0.04\;\; (0.07) \ , \end{eqnarray}
the bounds being at 90 (99\%) C.L. for 3 d.o.f. Unfortunately, it appears very
difficult to  probe (through present atmospheric data) values of $U^2_{e3}$ as
small as a few \%, which may entail interesting Earth matter effects
\cite{Fo99,Eart}. Constraining $U^2_{e3}$ is a major task for future
atmospheric \cite{Geis}, reactor \cite{Mika} and accelerator \cite{Acce} $\nu$
experiments.

The bounds on $3\nu$ mixing  are more evident  in the
$(\nu_e,\nu_\mu,\nu_\tau)$ triangle plot, embedding the unitarity constraint
(see \cite{Fo99,Fo97} for details). Figure~3 shows the allowed regions in such
triangle, whose lower and right sides represent the subcases of pure
$\nu_\mu\to\nu_\tau$ (allowed) and pure $\nu_\mu\to\nu_e$ (excluded). Large 
$\nu_e$ mixing is allowed by SK alone, but not by the SK+CHOOZ combination,
where only a narrow region survives near the lower side of the triangle.  In
such region, $U^2_{\mu3}\sim U^2_{\tau3}$ within a factor of two [e.g., 
$(U^2_{\mu3},\,U^2_{\tau3})\simeq(2/3,\,1/3)$ is also allowed]. In conclusion, the
$3\nu$ analysis of SK+CHOOZ shows that the $\nu_\mu\to\nu_e$ channel might be
open with a few $\%$ amplitude. Future atmospheric, reactor, and accelerator
$\nu$ experiments will test this interesting possibility.

\begin{figure*}
\vspace*{.5truecm}
\hspace*{1truecm}
\epsfig{bbllx=1.4truecm,bblly=1.3truecm,bburx=19.5truecm,bbury=26.5truecm,%
height=19truecm,figure=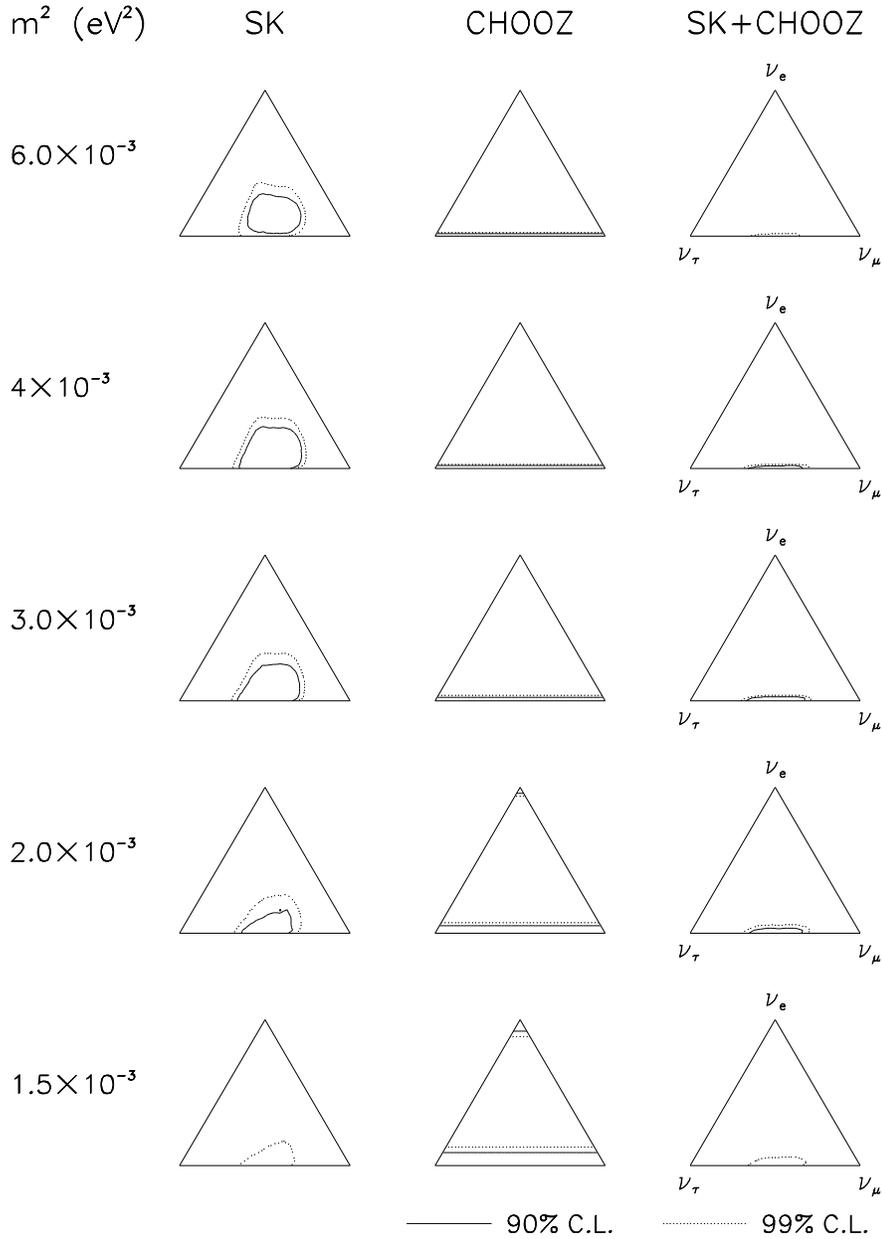}
\vspace*{-2.1cm}
\caption{Three-flavor analysis in the $(\nu_e,\nu_\mu,\nu_\tau)$ triangle plot,
for five representative values of $m^2$. Left and middle column: separate
analyses of Super-Kamiokande 70.5 kTy data and CHOOZ final data, respectively. 
Right column: combined SK+CHOOZ allowed regions. The SK+CHOOZ solutions are
close to pure $\nu_\mu\leftrightarrow\nu_\tau$ oscillations, with upper limits 
on $U^2_{e3}$ in the few percent range \protect\cite{Ma00}.}
\label{f3}
\end{figure*}

\section{$4\nu$ oscillations}

The current evidence for $\nu$ oscillations coming from  solar, atmospheric,
and LSND data can be accommodated by introducing a fourth, sterile neutrino
state $\nu_s$ \cite{Theo}. The  mass spectrum seems then to be favored in the
``2+2'' form (two separated doublets) \cite{2two}, although the ``3+1'' option
(triplet plus singlet) is not dismissed \cite{Fate}.

In 2+2 models, it is often assumed that atmospheric $\nu$ oscillations involve
{\em either\/} the $\nu_\mu\to\nu_\tau$ {\em or\/} the $\nu_\mu\to\nu_s$
channel. Correspondingly, it is assumed that  solar $\nu$ oscillations involve
{\em either\/}  the $\nu_e\to\nu_s$ {\em or\/} the $\nu_e\to\nu_\tau$ channel.
Such simplifying assumptions are challenged by the most recent SK data
\cite{SK00,Su00},  which disfavor oscillations into $\nu_s$ for both
atmospheric and solar neutrinos. However, it should be realized that
atmospheric $\nu_\mu$'s and solar $\nu_e$'s may also oscillate into {\em linear
combinations\/} of $\nu_s$ and $\nu_\tau$ \cite{Dool} (rather than into $\nu_s$
and $\nu_\tau$ separately), e.g., 
\begin{eqnarray}
{\rm atm.\ neutrino\ oscillations} &:& \nu_\mu\to\nu_+\ ,\\
{\rm solar\ neutrino\ oscillations} &:& \nu_e\to\nu_-\ ,
\end{eqnarray}
where
\begin{equation}
\left(\begin{array}{c}
\nu_+\\
\nu_-
\end{array}\right)=
\left(\begin{array}{cc}
+\cos\xi & +\sin\xi\\
-\sin\xi & +\cos\xi\\
\end{array}\right)
\left(\begin{array}{c}
\nu_\tau\\
\nu_s
\end{array}\right)\ ,
\end{equation}
with $\xi$ to be constrained by experiments.  A recent analysis of
$\nu_e\to\nu_-$ solar oscillations shows that all the usual solutions (MSW or
vacuum) are compatible with solar data for $\sin^2\xi > 0.3$ \cite{Conc}.

Concerning atmospheric $\nu$'s, we have analyzed \cite{Four} the same data as
in Fig.~2 for $\xi\in[0,\pi/2]$.  Figure~4 shows some representative results of
the $\chi^2$ fit, as a function of the mass square difference $m^2$. The fit
for unconstrained $\xi$ (thick solid line) is almost equal to the one for
$\xi=0$  (pure $\nu_\mu\to\nu_\tau$, thin solid line),  implying that the SK
data prefer small or zero admixture of $\nu_s$. The case $\xi=\pi/2$ (pure
$\nu_\mu\to\nu_s$, dashed line) leads to $\Delta\chi^2\simeq 15$ and is
disfavored. However, the case $\xi=\pi/4$ (fifty-fifty admixture of $\nu_\tau$
and $\nu_s$, dotted line) leads only to a modest increase in $\chi^2$ and
cannot be excluded.

The $4\nu$ analysis can also be done in a triangle plot (different from the
$3\nu$ case) embedding the  $(\nu_\mu,\nu_s,\nu_\tau)$  unitarity constraint
\cite{Four}. Figure~5 shows the results for separate and combined SK data sets.
It can be seen that, in the combination, the case of pure $\nu_\mu\to\nu_\tau$
oscillations (left side) is allowed, while the case of pure $\nu_\mu\to\nu_s$
oscillations (right side) is significantly disfavored. However, there are
intermediate solutions for $\sin^2\xi<0.7$ which have a significant admixture
of $\nu_s$. Such results and constraints emerge from the interplay of
low-energy data (which are more sensitive to $m^2$) and high-energy data (more
sensitive to the $\nu_s$ component through matter effects, scaling
as $\frac{1}{2}$~neutron density~$\times\sin^2\xi$ \cite{Four}).

\begin{figure}[t]
\vspace*{-.8truecm}
\hspace*{-0.7truecm}
\epsfig{bbllx=1.4truecm,bblly=1.3truecm,bburx=19.5truecm,bbury=26.5truecm,%
height=11.1truecm,figure=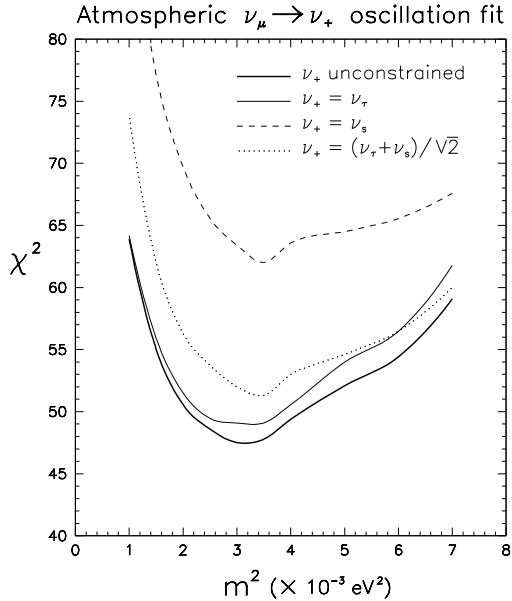}
\vspace*{-3.1cm}
\caption{ $\chi^2$ fit of 55 SK data bins (70.5 kTy)  for $\nu_\mu\to\nu_+$
oscillations, under various assumptions for the $\nu_s$ component of $\nu_+$. A
large  $\nu_s$ component (e.g., 50\%, dotted line)  is not excluded
\protect\cite{Four}.}
\label{f4}
\end{figure}

A qualitative comparison between such results \cite{Four} and those in 
\cite{Conc} indicates that atmospheric $\nu$ data can be reconciled with any of
the oscillation solutions to the solar $\nu$ problem in the range
$0.3<\sin^2\xi<0.7$. A somewhat different $4\nu$ analysis \cite{Yasu} derives
similar conclusions. Summarizing, it turns out that world $\nu$ oscillation
data are consistent with $4\nu$ solutions to the solar and atmospheric
anomalies, involving oscillations into both active and sterile states 
at the same time.

\begin{figure*}
\vspace*{.5truecm}
\hspace*{1truecm}
\epsfig{bbllx=1.4truecm,bblly=1.3truecm,bburx=19.5truecm,bbury=26.5truecm,%
height=19truecm,figure=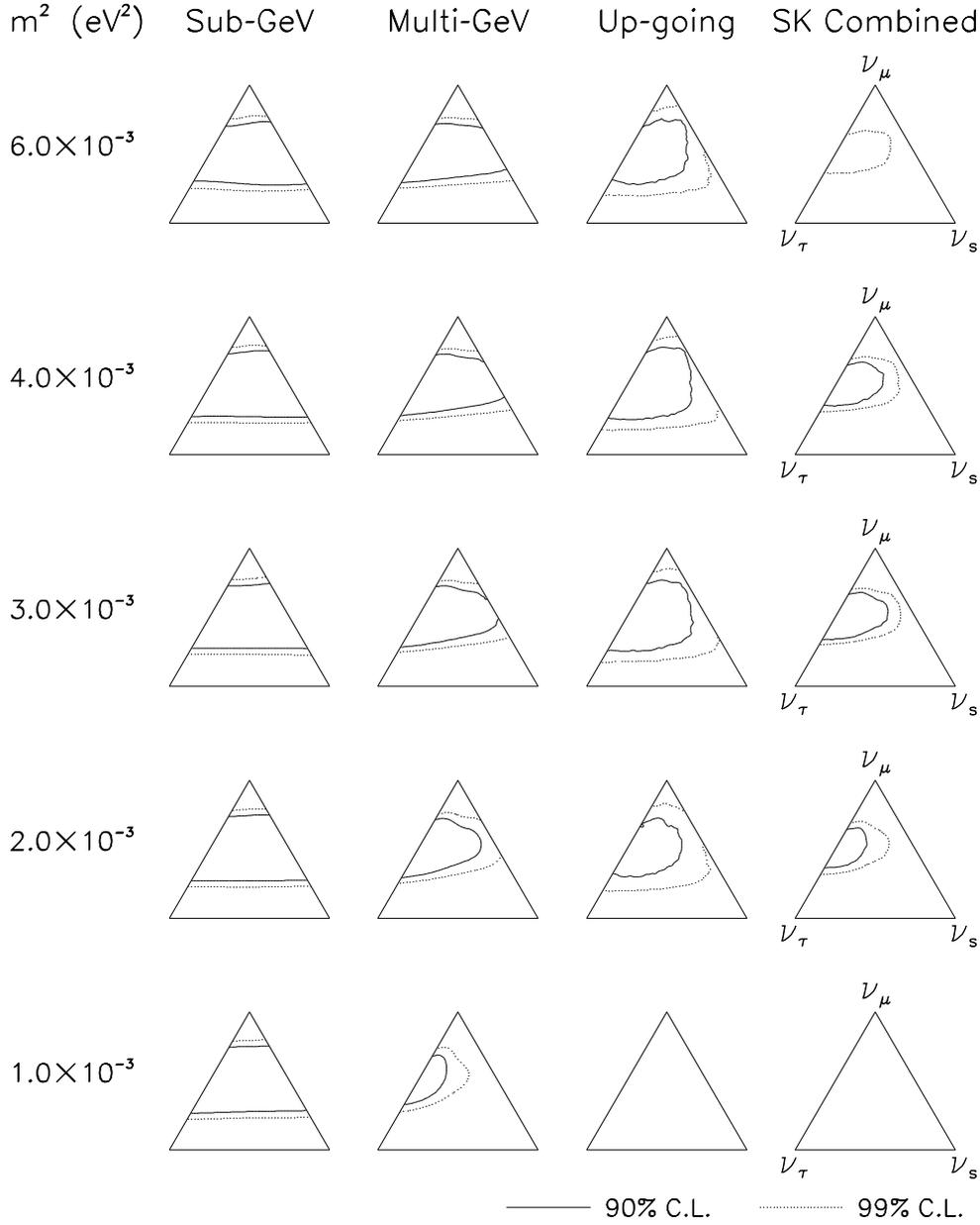}
\vspace*{-2.5cm}
\caption{$4\nu$ analysis in the $(\nu_\mu,\nu_s,\nu_\tau)$ triangle plot,  for
five representative values of $m^2$. First three columns: separate analyses of
SG$e$+SG$\mu$, MG$e$+MG$\mu$, and US$\mu$+UT$\mu$ data. Right column: all SK
data (70.5 kTy). The allowed regions  typically include pure
$\nu_\mu\leftrightarrow\nu_\tau$ oscillations (left side of the triangle) and
disfavor pure $\nu_\mu\to\nu_s$ oscillations (right side of the triangle).
However, intermediate situations with $\nu_\mu$ mixing with both $\nu_\tau$ and
$\nu_s$  are allowed inside the triangle \protect\cite{Four}.}
\label{f5}
\end{figure*}

\section{Nonstandard dynamics}

The SK data probe three decades in  pathlength $L$ and four decades in  energy
$E$. Such a wide  dynamical range severely constrains deviations from the
standard $L/E$ behavior of the $P_{\mu\tau}$  transition probability, which are
expected in the presence of exotic dynamics \cite{Lusi} (e.g., violations of
relativity principles \cite{Gasp,Glas}, which lead to a $L\cdot E$ behavior).

An analysis of older data (45 kTy) has shown that, assuming a $L\cdot E^n$
dependence of the phase, the SK measurements constrain $n$ to be very close to
$-1$, thus favoring standard oscillations, and excluding several nonstandard
explanations \cite{Expo}. Such results, shown in Fig.~6, have been strengthened
by the latest SK data \cite{SK00}. A peculiar FCNC scenario with $n=0$ 
\cite{FCNC} is also strongly disfavored---as any  energy independent mechanism
for $\nu_\mu$ disappearance---by combining low and high energy SK data
\cite{NOFC}. Therefore, $P_{\mu\tau}$ seems to be (dominantly) a function of
$L/E$.

However, is $P_{\mu\tau}$ necessarily a {\em periodic\/} function of $L/E$? The
answer is, surprisingly, no. There are (at least) three exotic scenarios which
predict a {\em monotonic\/} decrease of the oscillation probability in the
relevant $L/E$ range, and that are anyway reasonably consistent with the data.

\begin{figure}[bh]
\vspace*{-1.5truecm}
\hspace*{-0.6truecm}
\epsfig{bbllx=1.4truecm,bblly=1.3truecm,bburx=19.5truecm,bbury=26.5truecm,%
height=11.truecm,figure=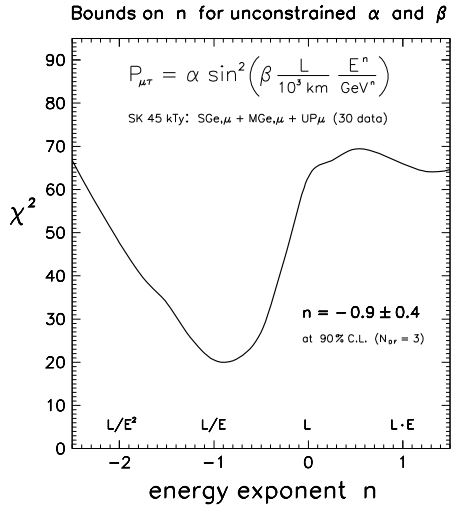}
\vspace*{-3.9cm}
\caption{ Bounds on the energy exponent $n$, assuming oscillation phase
$\propto L\cdot E^n$. (Older 45 kTy SK data used in this figure
\protect\cite{Expo}.)}
\label{f6}
\end{figure}

The first scenario involves $\nu$ decay \cite{Deca}, with a decay length of the
order of the Earth radius. The second scenario \cite{Barb} assumes $\nu_\mu$
mixing with neutrino states propagating in large extra dimensions \cite{Dien}.
A third scenario  \cite{Deco} assumes nonstandard Liouville dynamics
\cite{Bena}, leading to $\nu$ decoherence and thus to a damping of
oscillations. Figure~7 shows that the best fit for pure decoherence does not
differ significantly from the standard oscillation one \cite{Deco}.  The two
cases shown in Fig.~7 correspond to different functional forms for
$P_{\mu\mu}$,
\begin{eqnarray}
{\rm oscillation:\ }
P_{\mu\mu} &\simeq& \textstyle\frac{1}{2}[1+\cos(+\rho L/E)] 
\ ,\\
{\rm decohere. :\ }
P_{\mu\mu} &\simeq& \textstyle\frac{1}{2}[1+\exp(-\rho L/E)]\ ,
\end{eqnarray}
with $[E]=$~GeV, $[L]=$~km, and $\rho\simeq 7\times 10^{-3}$ GeV/km in both
cases. Such forms have the same asymptotic behavior, namely, $\langle
P_{\mu\mu}\rangle\simeq 1 (\frac{1}{2})$ for small (large) $L/E$, but they
significantly differ for intermediate values of $L/E$ where, however, the large
energy-angle smearing of SK prevents a clear discrimination.

Although  such nonstandard explanations \cite{Deca,Barb,Deco} of SK data do
not survive Occam's razor, they survive the current experimental tests for a
simple reason: the oscillation pattern (appearance of $\nu_\tau$ and {\em
re}-appearance of $\nu_\mu$) has not been directly observed so far, and a
monotonic $\nu_\mu$ disappearance is not excluded yet.  Therefore, the
unambigous observation of an oscillation cycle represents an important task
for future atmospheric \cite{Geis} and accelerator \cite{Acce} $\nu$
experiments.

\begin{figure}[hb]
\vspace*{-3.6truecm}
\hspace*{-0.5truecm}
\epsfig{bbllx=1.4truecm,bblly=1.3truecm,bburx=19.5truecm,bbury=26.5truecm,%
height=11.truecm,figure=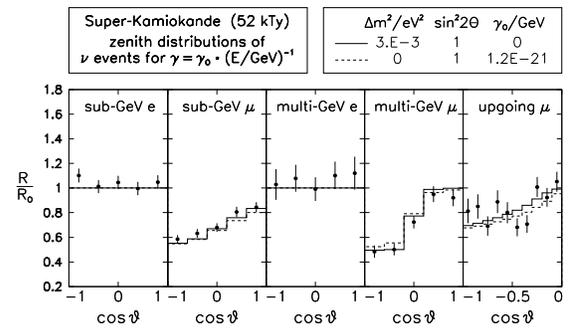}
\vspace*{-4.7cm}
\caption{Comparison of standard oscillations (solid line) 
{\em vs} neutrino decoherence (dashed line)
as explanation of the SK data. See \protect\cite{Deco} for details.}
\label{f7}
\end{figure}

\section{Conclusions}

Two-flavor $\nu_\mu\to\nu_\tau$ oscillations represent  a simple and beautiful
explanation of the SK data (as well as of MACRO and Soudan2). However one
cannot exclude, in addition,  subleading $\nu_\mu\to \nu_e$ transitions
(possible in $3\nu$ models) or sizable $\nu_\mu\to\nu_s$ transitions (possible
in $4\nu$ models). Moreover, the nonobservation of an oscillation cycle still
leaves room for exotic dynamics. Further experimental and theoretical work is
needed to firmly establish both the flavors and the dynamics involved in
atmospheric $\nu_\mu$ disappearance.


\end{document}